# Monolithic photonic chips for multi-channel frequency mixers and single photon detectors

Ming-Yang Zheng,[1] Quan Yao,[1] Bing Wang,[1,2,3] Xiuping Xie,[1] Qiang Zhang,[1,2,3,*] and Jian-Wei Pan,[2,3,*]

[1]Jinan Institute of Quantum Technology, Jinan 250101, China
[2]Shanghai Branch, CAS Center for Excellence and Synergetic Innovation Center in Quantum Information and Quantum Physics, University of Science and Technology of China, Shanghai 201315, China
[3]National Laboratory for Physical Sciences at Microscale and Department of Modern Physics, University of Science and Technology of China, Hefei 230026, China
*Corresponding author: qiangzhang@ustc.edu.cn, pan@ustc.edu.cn

**Abstract**

Lithium niobate photonic chip could realize diverse optical engineering for various applications benefiting from its excellent optical performances. In this letter, we demonstrate monolithic photonic chips for multi-channel sum-frequency conversion based on reverse-proton-exchange periodically poled lithium niobate waveguides, with the different channels showing uniform and excellent conversion efficiencies. To obtain a robust device and provide a convenient interface for applications, the integrated chip is fiber coupled with two fiber arrays. The packaged chip then forms the core of a multi-channel up conversion single photon detector. In each channel the input signal interacts with a 1950-nm single frequency pump laser and the sum frequency output is spectrally filtered and detected by a silicon avalanche photodiode. Average detection efficiency (DE) of 23.2 % and noise count rate (NCR) of 557 counts per second (cps) are achieved, with a standard deviation of 2.73 % and 48 cps over the 30 channels, as well as optical isolation (OI) between nearby channels of more than 71 dB, which are excellent for the extensive applications of monolithic photonic chips in fields including deep space laser communication, high-rate quantum key distribution and single-photon imaging.

## 1. Introduction

Photonic chips with multiple optical operations have extensive classical and quantum applications, such as optical-frequency measurement [1, 2], machine-learning [3], quantum communication and computing [4-6] and so on. As one of the building blocks for photonic circuits, frequency mixer has been investigated on several materials including AlN [7-9], GaAs [10, 11], lithium niobate [12-34], and others [35-40]. Owning to its wide transparency window (350–4500 nm), large optical nonlinearity efficient and flexible ferroelectric domain control, lithium niobate is considered as a promising integration platform.

Photonic chips for frequency mixing on lithium niobate have been extensively utilized for laser generation in visible and mid-infrared band by second-harmonic generation (SHG) and optical parametric oscillation (OPO) [12-14], entangled photon pairs generation by spontaneous parametric down conversion (SPDC) [15-19], quantum frequency converter by difference-frequency generation (DFG) [20-22], single photon detection by sum-frequency generation (SFG) [23-26], and others [27-34]. However, to the best of our knowledge, no work in literature has been focusing on multi-channel frequency mixing on a monolithic chip which could bring significant advances in the fields mentioned above with the features of system integration, low consumption, and low-cost. As a typical application, the multi-channel frequency mixers integrated on a monolithic chip could be employed to build multi-channel single photon detectors (SPDs) which have attracted much attention recently [41-43].

In this work, monolithic photonic chips based on reverse-proton-exchange (RPE) periodically poled lithium niobate (PPLN) waveguides are developed to construct a multi-channel up-conversion SPD in the telecommunication band. The photonic chip consisted of 34-channel sum-frequency conversion channels based on RPE PPLN waveguides is fiber coupled with two 34-channel fiber arrays. Test results show that the 30 adjacent channels in the middle have uniform and excellent conversion efficiencies. As the core device, the integrated chip is used to build a 30-channel up-conversion SPD, demonstrating an average DE of 23.2%, a NCR of 557 cps and an OI between nearby channels of more than 71 dB, which paves the way for its extensive applications in quantum information.

## 2. Waveguide fabrication and characterization

The multi-channel PPLN waveguide device is fabricated with the RPE technique which is a more mature technique [24, 25]. Photos of the packaged 34-channel integrated chip and the microscopic image of the waveguides are shown in Fig. 1. The 52mm*5.5mm multi-channel PPLN waveguide device cut from a 0.5 mm-thick wafer is composed of 34 adjacent independent channel waveguides with the same design parameters. The total length of each channel waveguide is 52 mm including mode filter, taper and the straight waveguide with quasi-phase-matching (QPM) gratings. A 1-mm-long 3.5 μm wide mode filter is located at the input port of the waveguide, followed by a 1-mm-long linear taper with the waveguide width increasing from 3.5 μm to 8.0 μm, with the latter kept through the remaining waveguide. The grating length is 48 mm with a poling period of 20 μm. A polarization-maintaining (PM) fiber array consisting of 34 pieces of 1550 nm PM single-mode fibers with a core spacing of 127 μm is used for input coupling. To ensure high and uniform coupling efficiencies between the PM fibers and the waveguides, the horizontal concentric axial spacing errors of the 34 fiber cores are required to be <0.5 μm. A multi-mode fiber array is used for output coupling, which is composed of 34-channel multi-mode fibers. Both the input and output end facets of the PPLN waveguide are anti-reflection coated to eliminate the Fresnel reflection loss.

A schematic diagram of the experimental setup for measuring the performances of the individual channels in the integrated device is shown in Fig. 2. The 1950-nm pump beam is combined with the 1550-nm signal via a 1550-nm/1950-nm wavelength division multiplexing (WDM) and then launched into a fiber pigtail with a microcapillary tube at the terminal. To achieve optimal coupling between the fiber and the PPLN waveguide, the microcapillary tube is mounted on a 6-axis manual stage for fine adjustment. The performances of the PPLN channel waveguides are measured one by one by translating the stage horizontally. The fibers in all the optical components including the lasers and WDM are polarization maintaining because the PPLN waveguide is fabricated with the RPE method and thus guide only the transverse-magnetic (TM) modes. The working temperature of the chip is controlled by a thermoelectric cooling (TEC) system to maintain the phase-matching condition.

The phase-matching wavelengths and conversion efficiencies of the 34-channel waveguides are measured respectively. The typical curves are shown in Fig. 3.

As shown in Fig. 3(a), the SFG tuning curve of the 16th channel waveguide is obtained by sweeping the signal wavelength around 1550 nm with the pump wavelength fixed at 1950 nm.

The phase-matching signal wavelength of the selected waveguide is 1550.4 nm at room temperature around 25 °C with a full width at half maximum (FWHM) of 0.65 nm.

The signal photon conversion efficiency η is

$$\eta = \frac{P_{SFG} \cdot \lambda_{SFG}}{P_{signal} \cdot \lambda_{signal}} \tag{1}$$

where $\lambda_{SFG}$ and $\lambda_{signal}$ are the SFG and signal wavelengths, $P_{SFG}$ is the SFG power at the output port of the waveguide, and $P_{signal}$ is the signal power at the input port of the waveguide which is fixed at 2 mW in the measurements.

Fig. 3(b) shows the conversion efficiencies versus the pump power measured at the output port of the 16th channel waveguide. As illustrated, the signal photon conversion efficiency of the waveguide reaches its maximum when the pump power is 96 mW, corresponding to a normalized conversion efficiency $\eta_{nor}$ of 116.3 %/(W·cm$^2$), which is calculated with [25]

$$\eta_{nor} = \frac{\pi^2}{4L^2 P_{max}} \tag{2}$$

where L is the length of the QPM gratings and $P_{max}$ is pump power required for maximum conversion.

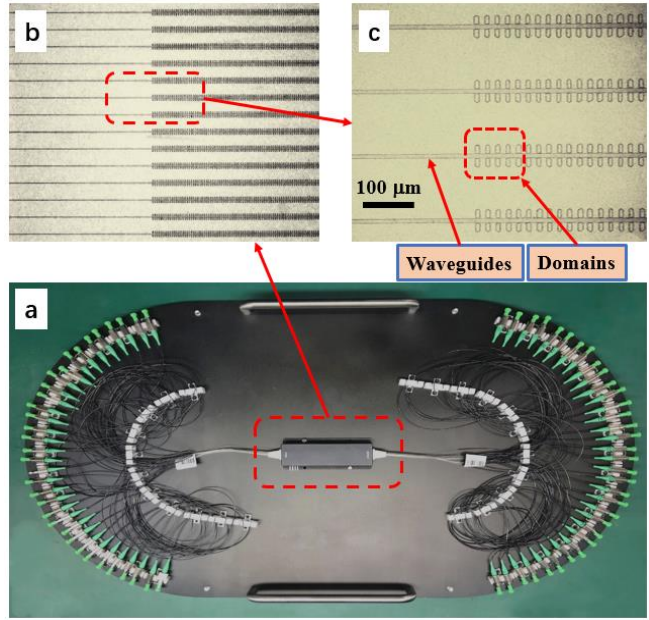

Fig. 1. (a) Photo of a packaged PPLN waveguide chip with 34-channel input and output fiber couplings; (b) Microscopic image of a multi-channel waveguide chip; (c) Details of the microstructure on the chip with a higher magnification of 200x. The chip shown in (b) and (c) is of the same design and processing as the packaged one in (a), and has been etched by hydrofluoric acid to reveal the waveguides and poling domains.

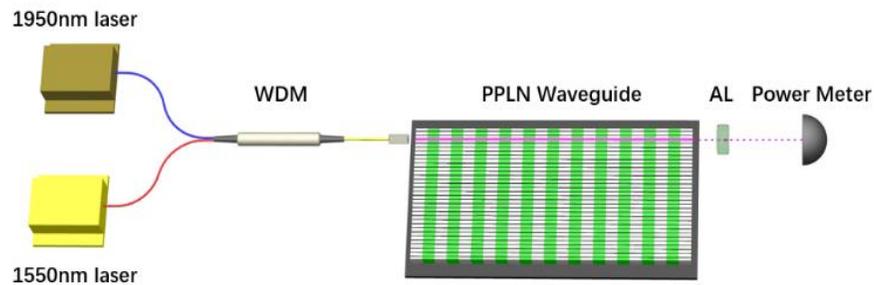

Fig. 2. Schematic diagram of the setup for measuring the performance of each channel in the integrated waveguide chip. WDM: wavelength division multiplexer, AL: aspherical lens.

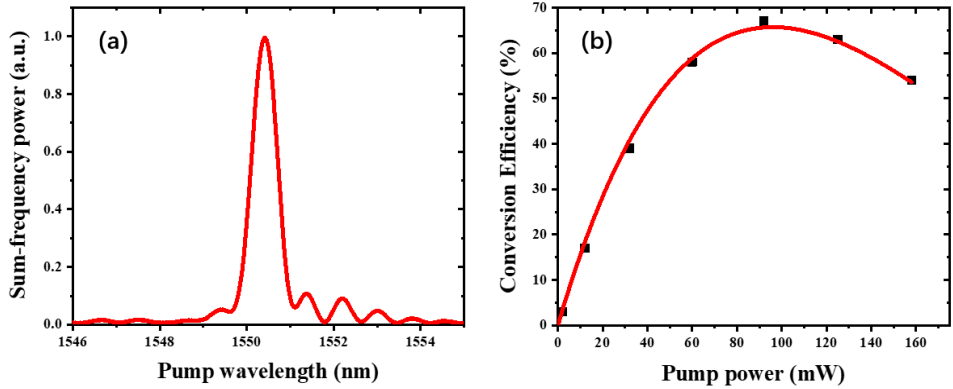

Fig. 3. (a) SFG tuning curve, (b) Measured SFG conversion efficiency versus pump power (solid square) and the sin²() fitting curve [24].

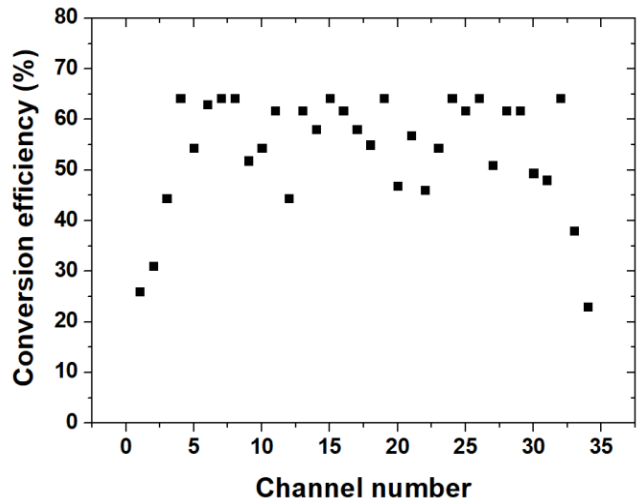

Fig. 4. Measured signal conversion efficiency (solid square) versus the waveguide channel number.

The signal photon conversion efficiency of each individual waveguide does not reach 100% due to the propagation loss in the waveguide and the coupling loss at the input port and other devices in the optical path. When the working temperature of the chip is kept at a certain temperature and the conversion efficiency is optimized for one channel waveguide by tuning the signal wavelength, the other channels may perform worse at that optimally tuned signal wavelength. In the following test, the signal and pump wavelengths are set at 1550 nm and 1950 nm respectively. An optimized working temperature at 23 °C is chosen for obtaining uniform performances over all the channels. The pump power at the maximum conversion efficiency point for each channel is a little different due to fabrication errors including the random duty-cycle errors of the QPM gratings and the waveguide width errors, and should be individually optimized accordingly when the integrated chip is used as a whole.

Calculated from the measured SFG power at the maximum conversion points of every channel, the maximum signal conversion efficiencies for all the 34 channels are shown in Fig. 4. We find that the conversion efficiencies of the first two and the last two channels are much lower than the other 30 channels in the middle. This is mainly attributed to the temperature gradient and liquid turbulence in the fabrication ovens for proton exchange and reverse proton exchange processes. The differences over the 30 channels in the middle may come from fabrication errors including the random duty-cycle errors of the QPM gratings, the waveguide width errors and the temperature gradient in the fabrication ovens. To increase the number of channels with uniform high conversion efficiencies, we may improve temperature uniformity of the fabrication ovens, or simply decrease the spacing between waveguide channels, if fiber array with smaller spacing is available. With a fiber to fiber (and waveguide center to center) spacing of 63 μm we may easily obtain 60 channels with uniform high conversion efficiencies.

## 3. Multi-channel up-conversion SPD

Here we show experiments with the 30-channel up-conversion SPD based on the packaged and integrated PPLN device. Note that the 4 side channels with much lower conversion efficiencies are excluded in the test here. Figure 5 shows the setup for single photon detection test [26], in which only one channel of the chip is employed for simplicity and clarity. A single-frequency fiber laser fixed at 1950 nm serves as the pump source. To correctly measure the detection efficiency, a single photon source (SPS) consisted of two variable optical attenuators (VOAs) and a 99/1 beam splitter is employed to provide one million PM photons near 1550 nm. A calibrated power meter (Keysight 81634B) is utilized to monitor the input signal power and ensures correct signal photon count. The SPS and the pump beam are combined with a 1550-nm/1950-nm WDM and launched into one channel of the fiber arrays at the input port of the PPLN waveguides. Employing PM fiber components for both the pump and the signal improves the stability of the whole system.

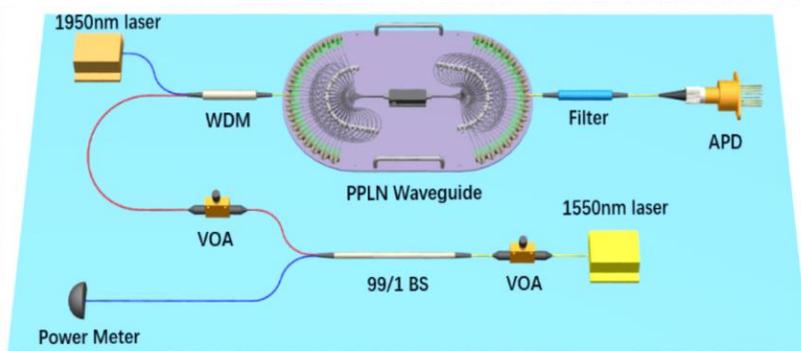

Fig. 5. Schematic of up-conversion SPD based on the 34-channel PPLN waveguide. VOA: variable optical attenuator, BS: beam splitter, APD: avalanche photodiode.

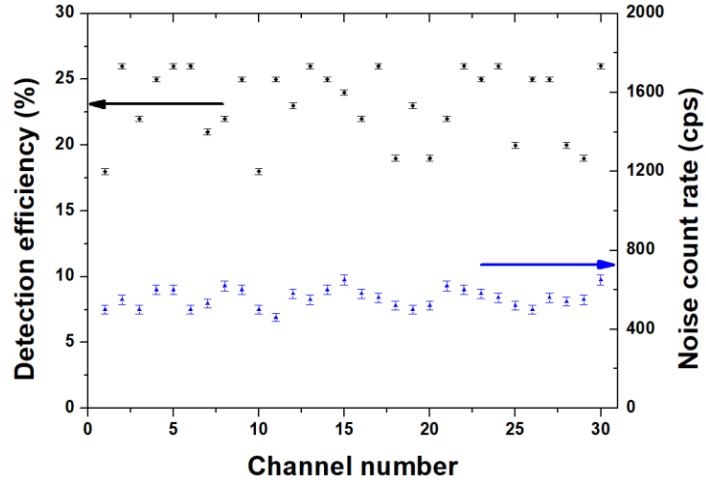

Fig. 6. Measured detection efficiency (black parallelogram) and dark count rate (blue triangle) versus the waveguide channel number. The error bars denote the shot noises. Note the similarity between the detection efficiency here and the conversion efficiency in Fig. 4, with a ratio of 40% between them caused by losses in the detection path.

To filter the noise photons generated from the pump laser, the up-conversion photons generated from the waveguide pass through a fiber filter. The fiber filter consists of an aperture and two band pass filters with a combined bandwidth of 0.5 nm. The noise photons mainly come from the spontaneous Raman scattering noise, parasitic noise caused by imperfect periodic poling structures, and second and third harmonic generation of the pump. The remaining photons are detected by a silicon APD with a detection efficiency of 55% and an intrinsic noise of 60 cps.

We tune the pump power of every channel to achieve maximum detection efficiency (DE). The system DE is obtained by dividing the number of detected count rate after NCR subtraction by one million, which is the signal photon count rate before entering the WDM. DEs and NCRs for all the channels are shown in Fig. 6. NCR contains the silicon APD's intrinsic noise.

The average DE and NCR of the middle 30 channels are 23.2 % and 557 cps, respectively. The standard deviations of the DE and NCR between different channels are 2.73 % and 48 cps respectively. DE and NCR of each channel can be tuned by the pump power, so we can achieve a complete consistency over all the 30 different channels by slightly adjusting the 1950-nm pump power in each channel with a VOA. The insertion losses of the WDM and the filter are 0.5 dB and 0.8 dB. The DEs measured here match with the data in Fig. 4 which is tested with milliwatt signal power when losses in the detection path are taken into account.

Optical isolation (OI) between nearby channels is important for a multi-channel single photon detector. To measure the OI, 1550-nm signal is launched into one channel with pump laser off, and the leakage signal coupled into an adjacent channel 127 μm away is up-converted with pump laser on and detected by Si APD. The measured OI between adjacent channels is >71 dB for a waveguide center to center spacing of 127 μm.

In section 2 we have proposed a device of 60 channels with waveguide center to center spacing of 63 μm. Without a fiber array with smaller core spacing for direct measurement of the SFG output, the OI for smaller channel spacing can be theoretically estimated with the isolation properties at the signal wavelength because the conversion efficiencies of the waveguides are uniform. Logarithm of the OI has a linear relation to the distance between two far away parallel waveguides [44], therefore the OI for waveguide center to center spacing of 63 μm can be deduced from those for waveguide center to center spacing of 50 μm and 127 μm which are available for measurement. From the measurements we estimate that the OI of the 60 channel single photon detector is >50 dB, much higher than the typical industrial standard of 30 dB. Limited by -30 dB crosstalk between adjacent channels, the maximum number of channels with a 5.5-mm-wide chip is ~120, which then needs a dense fiber array with core-core spacing of only 30 μm and is not commercially available yet.

## 4. Conclusions

In summary, a monolithic photonic chip for multi-channel sum-frequency conversion based on reverse-proton-exchange periodically poled lithium niobate waveguide is demonstrated. The input and output ports of the integrated chip are fiber coupled with two fiber arrays to realize a convenient interface and improve the device robustness. The packaged optical circuits show uniform and excellent conversion efficiencies for the different channels. Moreover, as a key component, the monolithic chip is used to build a multi-channel telecom-band up-conversion single photon detector. For the detector, an average detection efficiency of 23.2 % and a noise count rate of 557 counts per second are achieved, with a standard deviation of 2.73 % and 48 cps between 30 adjacent channels, as well as an ultra-high optical isolation between nearby channels of more than 71 dB. This detector may find immediate applications, such as free-space laser communications, quantum key distribution by wavelength division multiplexing, and the direct and non-line of sight single-photon imaging without scanning.

**Funding.**

National Key R&D Program of China (2018YFB0504300, 2017YFA0303902, 2017YFA0304000); Natural science foundation of Shandong province (ZR2019LLZ003); the

Chinese Academy of Science; the SAICT Experts Program; the Taishan Scholar Program of Shandong Province; Quancheng Industrial Experts Program; and the 5150 Program for Talents Introduction.

**Acknowledgements**

The authors thank Dai-Ying Wei for the packaging of the waveguides, Ji-Qian Yang, Jia Wang and Yang Gao for the help in the fabrication process.

**References**

[1]. D. T. Spencer, T. Drake, T. C. Briles, J. Stone, L. C. Sinclair, C. Fredrick, Q. Li, D. Westly, B. R. Ilic, A. Bluestone, N. Volet, T. Komljenovic, L. Chang, S. H. Lee, D. Y. Oh, M.-G. Suh, K. Y. Yang, M. H. P. Pfeiffer, T. J. Kippenberg, E. Norberg, L. Theogarajan, K. Vahala, N. R. Newbury, K. Srinivasan, J. E. Bowers, S. A. Diddams, and S. B. Papp, Nature 557, 81–85 (2018).

[2]. Z. L. Newman, V. Maurice, T. Drake, J. R. Stone, T. C. Briles, D. T. Spencer, C. Fredrick, Q. Li, D. Westly, B. R. Ilic, B. Shen, M. -G. Suh, K. Y. Yang, C. Johnson, D. M. S. Johnson, L. Hollberg, K. J. Vahala, K. Srinivasan, S. A. Diddams, J. Kitching, S. B. Papp, and M. T. Hummon, Optica 6, 680-685 (2019).

[3]. Y. Shen, N. C. Harris, S. Skirlo, M. Prabhu, T. Baehr-Jones, M. Hochberg, X. Sun, S. Zhao, H. Larochelle, D. Englund, and M. Soljačić, Nat. Photonics 11, 441 (2017).

[4]. X. Guo, C. -L. Zou, C. Schuck, H. Jung, R. Cheng, and H. X. Tang, Light Sci. Appl. 6, e16249 (2017).

[5]. S. Tanzilli, W. Tittel, M. Halder, O. Alibart, P. Baldi, N. Gisin, and H. Zbinden, Nature 437, 116–120 (2005).

[6]. T. D. Ladd, F. Jelezko, R. Laflamme, Y. Nakamura, C. Monroe, and J. L. O'Brien, Nature 464, 45–53 (2010).

[7]. X. Guo, C. -L. Zou, and H. X. Tang, Optica 3, 1126 (2016).

[8]. J. B. Surya, X. Guo, C. -L. Zou, and H. X. Tang, Opt. Lett. 43, 2696–2699 (2018).

[9]. A. W. Bruch, X. Liu, X. Guo, J. B. Surya, Z. Gong, L. Zhang, J. Wang, J. Yan, and H. X. Tang, Appl. Phys. Lett. 113, 131102 (2018).

[10]. L. Chang, A. Boes, X. Guo, D. T. Spencer, M. J. Kennedy, J. D. Peters, N. Volet, J. Chiles, A. Kowligy, N. Nader, D. D. Hickstein, E. J. Stanton, S. A. Diddams, S. B. Papp, and J. E. Bowers, Laser Photon. Rev. 12, 1800149, (2018).

[11]. P. S. Kuo, J. Bravo-Abad, and G. S. Solomon, Nat. Commun. 5, 3109 (2014).

[12]. K. -D. F. Büechter, H. Herrmann, C. Langrock, M. M. Fejer, and W. Sohler, Opt. Lett. 34, 470-472 (2009).

[13]. V. Mahal, A. Arie, M. A. Arbore, and M. M. Fejer, Opt. Lett. 21, 1217-1219 (1996).

[14]. C. Heese, C. R. Phillips, B. W. Mayer, L. Gallmann, M. M. Fejer, and U. Keller, Opt. Express 20, 26888-26894 (2012).

[15]. N. Montaut, L. Sansoni, E. Meyer-Scott, R. Ricken, V. Quiring, H. Herrmann, and C. Silberhorn, Phys. Rev. Applied 8, 024021 (2017).

[16]. C. -W. Sun, S. -H. Wu, J. -C. Duan, J. -W. Zhou, J. -L. Xia, P. Xu, Z. Xie, Y. -X. Gong, and S. -N. Zhu, Opt. Lett. 44, 5598-5601 (2019).

[17]. L. Yu, C. M. Natarajan, T. Horikiri, C. Langrock, J. S. Pelc, M. G. Tanner, E. Abe, S. Maier, C. Schneider, S. Höfling, M. Kamp, R. H. Hadfield, M. M. Fejer, and Y. Yamamoto, Nat. Commun. 6, 8955 (2015).

[18]. Y. Li, Y. Huang, T. Xiang, Y. Nie, M. Sang, L. Yuan, and X. Chen, Phys. Rev. Lett. 123, 250505 (2019).

[19]. T. Xiang, Y. Li, Y. Zheng, and X. Chen, Opt. Express 25, 12493-12498 (2017).

[20]. K. De Greve, L. Yu, P. L. McMahon, J. S. Pelc, C. M. Natarajan, N. Y. Kim, E. Abe, S. Maier, C. Schneider, M. Kamp, S. Höfling, R. H. Hadfield, A. Forchel, M. M. Fejer, and Y. Yamamoto, Nature 491, 421-425 (2012).

[21]. V. Esfandyarpour, C. Langrock, and M. Fejer, Opt. Lett. 43, 5655-5658 (2018).

[22]. Y. Yu, F. Ma, X. -Y Luo, B. Jing, P. -F. Sun, R. -Z. Fang, C. -W. Yang, H. Liu, M. -Y. Zheng, X. -P. Xie, W. -J. Zhang, L. -X. You, Z. Wang, T. -Y. Chen, Q. Zhang, X. -H. Bao, and J. -W. Pan, Nature 578, 240–245 (2020).

[23]. F. Ma, L. -Y. Liang, J. -P. Chen, Y. Gao, M. -Y. Zheng, X. -P. Xie, H. Liu, Q. Zhang, and J. -W. Pan, J. Opt. Soc. Am. B 35, 2096-2101 (2018).

[24]. J. S. Pelc, L. Ma, C. R. Phillips, Q. Zhang, C. Langrock, O. Slattery, X. Tang, and M. M. Fejer, Opt. Express 19, 21445-21456 (2011).

[25]. C. Langrock, E. Diamanti, R. V. Roussev, Y. Yamamoto, M. M. Fejer, and H. Takesue, Opt. Lett. 30, 1725-1727 (2005).

[26]. G. -L. Shentu, J. S. Pelc, X. -D. Wang, Q. -C. Sun, M. -Y. Zheng, M. M. Fejer, Q. Zhang, and J. -W. Pan, Opt. Express 21, 13986-13991 (2013).

[27]. J. Lin, N. Yao, Z. Hao, J. Zhang, W. Mao, M. Wang, W. Chu, R. Wu, Z. Fang, L. Qiao, W. Fang, F. Bo, and Y. Cheng, Phys. Rev. Lett. 122, 173903 (2019).

[28]. M. Jankowski, C. Langrock, B. Desiatov, A. Marandi, C. Wang, M. Zhang, C. R. Phillips, M. Lončar, and M. M. Fejer, Optica 7, 40-46 (2020).

[29]. J. Lu, J. B. Surya, X. Liu, A. W. Bruch, Z. Gong, Y. Xu, and H. X. Tang, Optica 6, 1455-1460 (2019).

[30]. R. Luo, Y. He, H. Liang, M. Li, and Q. Lin, Laser Photon. Rev. 13, 1800288 (2019).

[31]. R. Kou, S. Kurimura, K. Kikuchi, A. Terasaki, H. Nakajima, K. Kondou, and J. Ichikawa, Opt. Express 19, 11867-11872 (2011).

[32]. M. F. Volk, S. Suntsov, C. E. Rüter, and D. Kip, Opt. Express 24, 1386-1391 (2016).

[33]. M. Allgaier, V. Ansari, L. Sansoni, C. Eigner, V. Quiring, R. Ricken, G. Harder, B. Brecht, and C. Silberhorn, Nat. Commun. 8, 14288 (2017).

[34]. S. Tanzilli, A. Martin, F. Kaiser, M. P. De Micheli, O. Alibart, and D. B. Ostrowsky, Laser & Photonics Reviews 6, 115-143 (2011).

[35]. D. J. Wilson, K. Schneider, S. Hönl, M. Anderson, Y. Baumgartner, L. Czornomaz, T. J. Kippenberg, and P. Seidler, Nat. Photonics 14, 57-62 (2020).

[36]. X. Xue, F. Leo, Y. Xuan, J. A. Jaramillo-Villegas, P.-H. Wang, D. E. Leaird, M. Erkintalo, M. Qi, and A. M. Weiner, Light Sci. Appl. 6, e16253 (2017).

[37]. C. Xiong, W. Pernice, K. K. Ryu, C. Schuck, K. Y. Fong, T. Palacios, and H. X. Tang, Opt. Express 19, 10462–10470 (2011).

[38]. I. Roland, M. Gromovyi, Y. Zeng, M. El Kurdi, S. Sauvage, C. Brimont, T. Guillet, B. Gayral, F. Semond, J. Y. Duboz, M. de Micheli, X. Checoury, and P. Boucaud, Sci. Rep. 6, 34191 (2016).

[39]. K. Rivoire, Z. Lin, F. Hatami, W. T. Masselink, and J. Vučković, Opt. Express 17, 22609–22615 (2009).

[40]. S. Yamada, B. -S. Song, S. Jeon, J. Upham, Y. Tanaka, T. Asano, and S. Noda, Opt. Lett. 39, 1768–1771 (2014).

[41]. S. Miyajima, M. Yabuno, S. Miki, T. Yamashita, and H. Terai, Opt. Express 26, 29045-29054 (2018).

[42]. R. Cheng, X. Guo, X. Ma, L. Fan, K. Y. Fong, M. Poot, and H. X. Tang, Opt. Express 24, 27070-27076 (2016).

[43]. S. Miki, T. Yamashita, M. Fujiwara, M. Sasaki, and Z. Wang, Opt. Lett. 35, 2133-2135 (2010).

[44]. N. Ortega-Quijano, F. Fanjul-Vélez, and J. L. Arce-Diego, Opt. Commun., 283, 633-638 (2010).